\documentstyle[aps]{revtex}

\begin{document}
\draft
\title{Knotlike Cosmic Strings in The Early Universe}
\author{Yi-shi Duan and Xin Liu\thanks{%
Corresponding author. Electronic address: liuxin@lzu.edu.cn}}
\address{{\it Institute of Theoretical Physics, Lanzhou University,}\\
{\it Lanzhou 730000, P. R. China}}
\maketitle

\begin{abstract}
In this paper, the knotlike cosmic strings in the Riemann-Cartan space-time
of the early universe are discussed. It has been revealed that the cosmic
strings can just originate from the zero points of the complex scalar
quintessence field. In these strings we mainly study the knotlike
configurations. Based on the integral of Chern-Simons $3$-form a topological
invariant for knotlike cosmic strings is constructed, and it is shown that
this invariant is just the total sum of all the self-linking and linking
numbers of the knots family. Furthermore, it is also pointed out that this
invariant is preserved in the branch processes during the evolution of
cosmic strings.
\end{abstract}

\pacs{PACS number(s): 98.80.Cq, 02.40.-k, 11.15.-q}

\section{Introduction}

The recent measurements of the redshift of type Ia supernova (SN Ia) and the
power spectrum of the cosmic microwave background (CMB) from BOOMERANG-98
and MAXIMA-1 respectively suggest the accelerating expansion and the
flatness of the universe \cite{2observation,Turner}. These two observations
lead to the conclusion that the universe has the critical density and
consists of $1/3$ of ordinary matter and $2/3$ of dark energy with negative
pressure. At the moment, a most-often considered candidate for dark energy
is the {\it quintessence} \cite{quintessence}. In the present paper, we
mainly consider the complex scalar quintessence field $\psi (x)$ on the
Riemann-Cartan manifold (${\bf U}^4$) of the early universe \cite{cmplxsclr}%
. As a background field of the universe, $\psi (x)$ is a section of the
complex line bundle, i.e., a section of the two-dimensional real vector
bundle on ${\bf U}^4$: 
\begin{equation}
\psi (x)=\phi ^1{\bf (}x)+i\phi ^2{\bf (}x).  \label{psiphi12}
\end{equation}
In this paper, in the following Sect.II, it will be shown that the cosmic
string structures can just originate from the zero points of this complex
scalar quintessence field $\psi (x)$.

In 1997, Faddeev and Niemi pointed out that for a string structure of finite
energy, its length must be finite, which is possible if its core forms a
knot. It is shown that in a realistic $(3+1)$-dimensional model there exist
knotlike structures appearing as stable finite solitons \cite{FadNie}. Since
then knotlike configurations as string structures of finite energy are paid
close attention to in a variety of physical, chemical and biological
scenarios, including the structure of elementary particles \cite{knotpart},
the early universe cosmology \cite{Schiappa,collision}, the Bose-Einstein
condensation \cite{knotspcond}, the polymer folding \cite{knotpoly}, and the
DNA replication, transcription and recombination \cite{knotbio}. In the
following Sect.III, we will emphasize on the knotlike cosmic strings and
study the topological invariant for these knots.

This paper is arranged as follows. In Sect.II, we use the $\phi $-mapping
topological current theory \cite{topcurrent2,topcurrent} to reveal that the
cosmic strings can just originate from the zero points of quintessence field 
$\psi (x)$. In this section the Nielsen Lagrangian and Nambu action of these
strings are also simply discussed. In Sect.III, we mainly study the knotlike
configurations in these string structures. Based on the integral of
Chern-Simons $3$-form we construct a topological invariant for the knotlike
strings. It is shown that this invariant is just the total sum of all the
self-linking and linking numbers of the knots family. In Sect.IV, moreover,
the conservation of this topological invariant in the branch processes (i.e.
the splitting, mergence and intersection) during the evolution of knotlike
cosmic strings is simply discussed.

\section{The Cosmic Strings}

In this section, it is pointed out that the cosmic strings can originate
from the zero points of the complex scalar quintessence field $\psi (x)$.
The Nielsen Lagrangian and Nambu action of these strings are also simply
discussed.

The $U(1)$ gauge field tensor, i.e. the curvature of $U(1)$ principal bundle
on ${\bf U}^4$, is written as 
\begin{equation}
F_{\mu \nu }=\partial _\mu A_\nu -\partial _\nu A_\mu ,  \label{TuvU1}
\end{equation}
where $\mu =0,1,2,3$ denotes the base manifold, and $A_\mu $ is the $U(1)$
gauge potential, i.e. the connection of $U(1)$ principal bundle on ${\bf U}%
^4 $. Define $A_\mu $ possesses the inner structure as 
\begin{equation}
A_\mu =\frac \alpha {2\pi }\frac 1{2i\psi ^{*}\psi }(\psi ^{*}\partial _\mu
\psi -\partial _\mu \psi ^{*}\psi ),  \label{AudecomU1}
\end{equation}
where $\alpha =\sqrt{\hbar G/c^3}$ is a constant. It can be proved that
under the $U(1)$ gauge transformation of $\psi (x)$: $\psi ^{\prime
}=e^{i\lambda }\psi ,$ $A_\mu $ satisfies the $U(1)$ transformation
relation: 
\begin{equation}
A_\mu ^{\prime }=A_\mu +i\frac \alpha {2\pi }\partial _\mu \lambda ,
\label{Autrans}
\end{equation}
where $\lambda \in {\bf R}$\ is the transformation parameter. [In (\ref
{AudecomU1}) the {\it RHS }has taken the form of velocity field in quantum
mechanics. In superconductivity theory, the form of (\ref{AudecomU1})
actually corresponds to the London relation \cite{London}].

Introducing the two-dimensional unit vector $n^a$ from $\phi ^{1,2}$: 
\begin{equation}
n^a=\frac{\phi ^a}{\left\| \phi \right\| },\;\;\;\;(a=1,2;\;\left\| \phi
\right\| ^2=\phi ^a\phi ^a=\psi ^{*}\psi )
\end{equation}
the expression (\ref{AudecomU1}) can be written as 
\begin{equation}
A_\mu =\frac \alpha {2\pi }\epsilon _{ab}n^a\partial _\mu n^b,
\label{AudecomU1'}
\end{equation}
and then the field tensor $F_{\mu \nu }$ is just 
\begin{equation}
F_{\mu \nu }=\frac \alpha {2\pi }2\epsilon _{ab}\partial _\mu n^a\partial
_\nu n^b.  \label{Tuvnanb}
\end{equation}
According to the $\phi $-mapping topological current theory \cite
{topcurrent2}, in Riemann-Cartan space-time the topological tensor current
is defined as 
\begin{equation}
j^{\mu \nu }=\frac 12\frac 1{\sqrt{g}}\epsilon ^{\mu \nu \lambda \rho
}F_{\lambda \rho }.\;\;(g=\det \{g_{\mu \nu }\})  \label{juv}
\end{equation}
It is seen that (\ref{juv}) satisfies 
\begin{equation}
j^{\mu \nu }=-j^{\nu \mu },\;\;\;\nabla _\mu j^{\mu \nu }=\frac 1{\sqrt{g}}%
\partial _\mu (\sqrt{g}j^{\mu \nu })=0,
\end{equation}
hence $j^{\mu \nu }$ is anti-symmetric and is an identically conserved
current.

Using $\partial _\mu n^a=\frac{\partial _\mu \phi ^a}{\Vert \phi \Vert }%
+\phi ^a\partial _\mu \frac 1{\Vert \phi \Vert }$ {and the Green function
relation in }$\phi $-{space: }${\partial }_a{\partial }_a{\ln \Vert \phi
\Vert =2\pi \delta ^2(\vec{\phi})\;}$(where ${{\partial }_a{=\partial
/\partial }\phi ^a}$), it can be proved that $j^{\mu \nu }$ may be expressed
in a $\delta $-function form: 
\begin{equation}
j^{\mu \nu }=\frac \alpha {\sqrt{g}}\delta ^2(\vec{\phi})D^{\mu \nu }(\frac %
\phi x),  \label{deltav}
\end{equation}
where{\ }${D^{\mu \nu }(\phi /x)=\frac 12\epsilon }^{\mu \nu \rho \lambda }{%
\epsilon _{ab}\partial }_\rho {\phi }^a{\partial _\lambda \phi }^b$; while
the spatial component of $j^{\mu \nu }$ is just 
\begin{equation}
j^i=j^{0i}=\frac 1{2\sqrt{g}}\epsilon ^{ijk}F_{jk}=\frac \alpha {\sqrt{g}}%
\delta ^2(\vec{\phi})D^i(\frac \phi x),\;\;\;(i,j,k=1,2,3)  \label{jidel}
\end{equation}
where $D^i(\phi /x)=\frac 12{\epsilon }^{ijk}{\epsilon _{ab}\partial }_j{%
\phi }^a{\partial _k\phi }^b$ is the Jacobian vector.

Obviously the expression (\ref{deltav}) provides an important conclusion: 
\begin{equation}
j^{\mu \nu }\left\{ 
\begin{array}{l}
=0,\;iff\;\vec{\phi}\neq 0; \\ 
\neq 0,\;iff\;\vec{\phi}=0,
\end{array}
\right.
\end{equation}
so it is necessary to study the zero points of $\vec{\phi}$ to determine the
non-zero solutions of $j^{\mu \nu }$. The implicit function theory shows 
\cite{Goursat} that under the regular condition 
\begin{equation}
D^{\mu \nu }(\phi /x)\neq 0,  \label{regulcond}
\end{equation}
the general solutions of 
\begin{equation}
\phi ^1(x^0=t,\;x^1,\;x^2,\;x^3)=0,\;\phi ^2(x^0=t,\;x^1,\;x^2,\;x^3)=0
\end{equation}
can be expressed as 
\begin{equation}
x^\mu =x_k^\mu (u^1,\;u^2),\;\;\;(\mu =0,1,2,3)  \label{paramet}
\end{equation}
which represents $N$ two-dimensional singular submanifolds $P_k$ $%
(k=1,2,...,N)$ with intrinsic coordinates $u^1$ and $u^2$. In this paper, in
particular, $u^1$ and $u^2$ respectively take the space-like string
parameter $s$ and time-like evolution parameter $t$ (i.e. $u^1=s,\;u^2=t$),
then the $N$ submanifolds $P_k$'s are just the world surfaces of a family of 
$N$ moving isolated singular strings $L_k$'s. These singular string
solutions are just the cosmic strings.

Next we should expand $j^{\mu \nu }$ onto these $N$ singular submanifolds $%
P_k$'s. First, it can be proved that in the four-dimensional space-time
there exists a two-dimensional submanifold $M$ which is transversal to every 
$P_k$ at the section point $p_k$: 
\begin{equation}
\left. g_{\mu \nu }\frac{\partial x^\mu }{\partial u^I}\frac{\partial x^\nu 
}{\partial v^C}\right| _{p_k}=0,\;\;\;(I=1,2;\;C=1,2)
\end{equation}
where $v^1$ and $v^2$ are the intrinsic coordinates of $M$. Then on $M$ one
can prove that \cite{Schouten,topcurrent2} 
\begin{equation}
\delta ^2(\vec{\phi})=\sum_{k=1}^N\beta _k\eta _k\delta ^2(\vec{v}-\vec{v}%
(p_k)),  \label{deltv}
\end{equation}
where the positive integer $\beta _k$ is the Hopf index and $\eta _k=\pm 1$
the Brouwer degree of $\phi $-mapping, and $W_k=\beta _k\eta _k$ is just the
winding number of string $L_k$. Second, since every $p_k$ is related to a
singular submanifold $P_k$, the above two-dimensional $\delta $-function of
singular point [i.e. $\delta ^2(\vec{v}-\vec{v}(p_k))$] must be expanded to
the $\delta $-function on singular submanifold $P_k$ [i.e. $\delta (P_k)$].
Meanwhile in $\delta $-function theory it has been given that \cite
{Schouten,topcurrent2} 
\begin{equation}
\delta (P_k)=\int_{P_k}\delta ^4(x^\mu -x_k^\mu (u))\sqrt{g_u}d^2u,
\label{deltNk}
\end{equation}
where $g_u$ is the determinant of the metric $g_{IJ}$ of $P_k$: $g_u=\det
(g_{IJ})\;(I,J=1,2;\;g_{IJ}=g_{\mu \nu }\frac{\partial x^\mu }{\partial u^I}%
\frac{\partial x^\nu }{\partial u^J}).$ So, third, from (\ref{deltNk}) and (%
\ref{deltv}), we have 
\begin{equation}
\delta ^2(\vec{\phi})=\sum_{k=1}^N\beta _k\eta _k\int_{P_k}\delta ^4(x^\mu
-x_k^\mu (u))\sqrt{g_u}d^2u,  \label{deltsub}
\end{equation}
and therefore $j^{\mu \nu }$ is expanded onto $N$ singular submanifolds $P_k$%
: 
\begin{equation}
j^{\mu \nu }=\frac \alpha {\sqrt{g}}D^{\mu \nu }(\frac \phi x%
)\sum_{k=1}^N\beta _k\eta _k\int_{P_k}\delta ^4(x^\mu -x_k^\mu (u))\sqrt{g_u}%
d^2u.  \label{juvsubdelt}
\end{equation}
In (\ref{juvsubdelt}) the spatial components of $j^{\mu \nu }$ is 
\begin{equation}
j^i=j^{0i}=\frac 1{2\sqrt{g}}\epsilon ^{ijk}F_{jk}=\frac \alpha {\sqrt{g}}%
\sum_{k=1}^NW_k\int_{L_k}\frac{dx^i}{ds}\delta ^3(\vec{x}-\vec{x}%
_k(s))ds,\;\;\;(W_k=\beta _k\eta _k)  \label{vstr}
\end{equation}
where $\frac{dx^i}{ds}=\frac{D^i(\phi /x)}{D(\phi /u)}.$ Hence the
topological charge of string $L_k$ is just 
\begin{equation}
Q_k=\frac 1\alpha \int_{\Sigma _k}j^i\sqrt{g}d\sigma _i=W_k,
\end{equation}
where $\Sigma _k$ is the two-dimensional spatial surface element
perpendicular to $L_k$.

In the end of this section we would also simply discuss the Nambu action of
these $N$ cosmic strings $L_k$'s. Define the Lagrangian of these strings as 
\begin{equation}
L=\frac 1\alpha \sqrt{\frac 12g_{\mu \lambda }g_{\nu \rho }j^{\mu \nu
}j^{\lambda \rho }},
\end{equation}
which is just the generalization of Nielsen's Lagrangian \cite{Nielsen}.
Then from the above deduction (\ref{deltav}) one can prove 
\begin{equation}
L=\frac 1{\sqrt{g}}\delta (\vec{\phi}).
\end{equation}
So the action is just 
\begin{equation}
S=\int_{{\bf U}^4}L\sqrt{g}d^4x=\int_{{\bf U}^4}\delta (\vec{\phi})d^4x.
\label{2}
\end{equation}
Substituting (\ref{deltsub}) into (\ref{2}): 
\begin{eqnarray}
S &=&\int_{{\bf U}^4}\sum_{k=1}^N\beta _k\eta _k\int_{P_k}\delta ^4(x-x_k(u))%
\sqrt{g_u}d^2ud^4x  \nonumber \\
&=&\sum_{k=1}^N\beta _k\eta _k\int_{P_k}\sqrt{g_u}d^2u,
\end{eqnarray}
we arrive at an important result 
\begin{equation}
S=\sum_{k=1}^N\beta _k\eta _kS_k,  \label{Nambuact}
\end{equation}
where $S_k=\int_{P_k}\sqrt{g_u}d^2u$ is the area of singular submanifold $%
P_k $. Therefore the expression (\ref{Nambuact}) is just the Nambu action of
the $N$ cosmic strings \cite{Nielsen,Schiappa}, which is the basis of the
further work on cosmic string theory.

Furthermore, from the principle of least action, we can also obtain the
evolution equation for these $N$ strings as \cite{topcurrent2} 
\begin{equation}
\frac 1{\sqrt{g_u}}\frac \partial {\partial u^I}(\sqrt{g_u}g^{IJ}\frac{%
\partial x^\mu }{\partial u^J})+g^{IJ}\Gamma _{\nu \lambda }^\mu \frac{%
\partial x^\nu }{\partial u^I}\frac{\partial x^\lambda }{\partial u^J}%
=0.\;\;\;(I,J=1,2)
\end{equation}

Finally, it should be addressed that in the above text the regular condition
(\ref{regulcond}) has been used; when this condition fails, the branch
processes during the evolution of cosmic strings will occur. This will be
detailed in Sect.IV.

\section{Topological Invariant for Knotlike Cosmic Strings}

In this section, based on the integral of Chern-Simons $3$-form we mainly
study the topological invariant for the knotlike cosmic strings.

In order to construct a topological invariant in the space-time, one must
pick an integral expression which does not require any choice of metric $%
g_{\mu \nu }$. Precisely in three-dimensional space there is a reasonable
choice, namely, the integral of the Chern-Simons $3$-form \cite
{CSterm,Witten}: 
\begin{equation}
Q=(\frac{2\pi }\alpha )^2\frac 1{4\pi }\int_\Omega \epsilon
^{ijk}A_iF_{jk}d^3x,  \label{intCSform}
\end{equation}
where $\Omega $ is the spatial volume. Hereinafter we just study (\ref
{intCSform}) to get the topological invariant for the knotlike cosmic
strings.

Using the above (\ref{jidel}) and (\ref{vstr}), the expression (\ref
{intCSform}) can be written as 
\begin{equation}
Q=\frac{2\pi }\alpha \int_\Omega A_i\delta ^2(\vec{\phi})D^i(\frac \phi x%
)d^3x=\frac{2\pi }\alpha \sum_{k=1}^NW_k\int_{L_k}A_idx^i.  \label{CS1}
\end{equation}
It can be seen that when the $N$ cosmic strings of (\ref{CS1}) are $N$
closed curves, i.e., a family of $N$ knotlike strings $\gamma
_k\;(k=1,...,N) $, (\ref{CS1}) leads to 
\begin{equation}
Q=\frac{2\pi }\alpha \sum_{k=1}^NW_k\oint_{\gamma _k}A_idx^i.  \label{CSact3}
\end{equation}
This is a very important expression. Considering the $U(1)$ gauge
transformation of $A_i$ in (\ref{Autrans}): $A_i^{\prime }=A_i+i\frac \alpha
{2\pi }\partial _i\lambda ,$ it can be seen that the $(i\frac \alpha {2\pi }%
\partial _i\lambda )$ term contributes nothing to the integral $Q,$ hence
the expression (\ref{CSact3}) is invariant under the gauge transformation.
Therefore, from the fact that $Q$ of (\ref{CSact3}) is independent of the
choice of metric and is invariant under the $U(1)$ gauge transformation, one
can conclude that $Q$ is a topological invariant for the knotlike cosmic
strings. This can be used in the research of the topology of string
structures in the early universe. At the same time, for the Chern-Simons
integral itself, (\ref{CSact3}) provides a sufficiency condition for the
Chern-Simons integral to be a topological invariant, which is another
significance of expression (\ref{CSact3}).

In following we will show that $Q$ is just the total sum of all the
self-linking and linking numbers of the knotlike strings family. Using (\ref
{jidel}), the expression (\ref{CSact3}) can be reexpressed as 
\begin{equation}
Q=\frac{2\pi }\alpha \sum_{k=1}^N\sum_{l=1}^NW_kW_l\oint_{\gamma
_k}\oint_{\gamma _l}\partial _iA_jdx^idy^j,  \label{CSact4}
\end{equation}
where $\vec{x}$ and $\vec{y}$ are two points respectively on knots $\gamma
_k $ and $\gamma _l$. Noticing that $\gamma _k$ and $\gamma _l$ can be the
same one knot, or two different knots, we should write (\ref{CSact4}) in two
parts ($k=l$ and $k\neq l$); furthermore the $k=l$ part includes both the $%
\vec{x}\neq \vec{y}$ and the $\vec{x}=\vec{y}$ cases. So totally $Q$ should
be written in three terms: $(k=l;\;\vec{x}\neq \vec{y}),$ $(k=l$ $;$ $\vec{x}%
=\vec{y})$ and $(k\neq l)$.

Define a three-dimensional unit vector 
\begin{equation}
\vec{m}=\frac{\vec{y}-\vec{x}}{\left\| \vec{y}-\vec{x}\right\| },
\end{equation}
and another two-dimensional unit vector $\vec{e}$ on the $\vec{m}$%
-formed-sphere $S^2:$%
\begin{equation}
\vec{e}\bot \vec{m},\;\vec{e}\cdot \vec{e}=1.
\end{equation}
The vector field $\vec{e}$ is the section of two-dimensional real vector
bundle, i.e., the section of complex line bundle 
\begin{equation}
\chi =e^1+ie^2,
\end{equation}
where $\chi $ is the complex scalar field. Then, similarly as in Sect.II,
one can give out the inner structure of $A_i$ in terms of $\vec{e}$ (i.e. $%
\chi $) as 
\begin{equation}
A_i=\frac \alpha {2\pi }\epsilon _{ab}e^a\partial
_ie^b.\;\;\;\;(a,\;b=1,2;\;i=1,2,3)  \label{decomsec}
\end{equation}

Using (\ref{decomsec}) and the relation $2\epsilon _{ab}\partial
_ie^a\partial _je^b=\vec{m}\cdot (\partial _i\vec{m}\times \partial _j\vec{m}%
)$, the three terms of $Q$ can be expressed as 
\begin{eqnarray}
Q &=&2\pi [\sum_{k=1\;(\vec{x}\neq \vec{y})}^N\frac 1{4\pi }%
W_k^2\oint_{\gamma _k}\oint_{\gamma _k}\vec{m}^{*}(dS)+\frac 1{2\pi }%
\sum_{k=1}^NW_k^2\oint_{\gamma _k}\epsilon _{ab}e^a\partial _ie^bdx^i 
\nonumber \\
&&+\sum_{k,l=1\;(k\neq l)}^N\frac 1{4\pi }W_kW_l\oint_{\gamma
_k}\oint_{\gamma _l}\vec{m}^{*}(dS)].  \label{CSact8}
\end{eqnarray}
where $\vec{m}^{*}(dS)=$ $\vec{m}\cdot (\partial _i\vec{m}\times \partial _j%
\vec{m})dx^i\wedge dy^j\;(\vec{x}\neq \vec{y})$ denotes the pull-back of $%
S^2 $ surface element.

Let us discuss these three terms in detail. First, the first term of (\ref
{CSact8}) is just related to the writhing number $Wr(\gamma _k)$ of $\gamma
_k$ \cite{White}: 
\begin{equation}
Wr(\gamma _k)=\frac 1{4\pi }\oint_{\gamma _k}\oint_{\gamma _k}\vec{m}%
^{*}(dS).  \label{wrnum}
\end{equation}
For the second term of (\ref{CSact8}), since this is the $\vec{x}=\vec{y}$
term, one can prove that it is related to the twisting number $Tw(\gamma _k)$
of $\gamma _k$ \cite{White} 
\begin{equation}
\frac 1{2\pi }\oint_{\gamma _k}\epsilon _{ab}e^a\partial _ie^bdx^i=\frac 1{%
2\pi }\oint_{\gamma _k}(\vec{T}\times \vec{V})\cdot d\vec{V}=Tw(\gamma _k),
\label{twnum}
\end{equation}
where $\vec{T}$ is the unit tangent vector of knot $\gamma _k$ at $\vec{x}$ [%
$\vec{m}=\vec{T}$ when $\vec{x}=\vec{y}$], and $\vec{V}$ is defined as $%
e^a=\epsilon ^{ab}V^b\;(\vec{V}\bot \vec{T},\;\vec{e}=\vec{T}\times \vec{V})$%
. From the White formula \cite{White} 
\begin{equation}
SL(\gamma _k)=Wr(\gamma _k)+Tw(\gamma _k)  \label{white}
\end{equation}
one see that the first and second terms of (\ref{CSact8}) just compose the
self-linking numbers of knots.

Second, for the third term, one can prove 
\begin{equation}
\frac 1{4\pi }\oint_{\gamma _k}\oint_{\gamma _l}\vec{m}^{*}(dS)=\frac 1{4\pi 
}\epsilon ^{ijk}\oint_{\gamma _k}dx^i\oint_{\gamma _l}dy^j\frac{(x^k-y^k)}{%
\left\| \vec{x}-\vec{y}\right\| ^3}=Lk(\gamma _k,\gamma _l)\;\;(k\neq l)
\label{Gausslink}
\end{equation}
where $Lk(\gamma _k,\gamma _l)$ is the Gauss linking number between $\gamma
_k$ and $\gamma _l$\cite{Witten,Polyakov}.

Therefore, third, from (\ref{wrnum}), (\ref{twnum}), (\ref{white}) and (\ref
{Gausslink}), we arrive at the important result: 
\begin{equation}
Q=2\pi [\sum_{k=1}^NW_k^2SL(\gamma _k)+\sum_{k,l=1\;(k\neq
l)}^NW_kW_lLk(\gamma _k,\gamma _l)].  \label{CSact9}
\end{equation}
This precise expression just reveals the relationship between $Q$ and the
self-linking and linking numbers of the knots family \cite
{Witten,Bekenstein,Kozhevnikov}. Since the self-linking and linking numbers
are both the intrinsic invariant characteristic numbers of knots family in
topology, expression (\ref{CSact9}) directly relates $Q$ to the topology of
the knots family itself, and therefore $Q$ can be regarded as an important
invariant required to describe the topology of knotlike cosmic strings in
early universe. This is just the significance of the introduction and
research of topological invariant $Q$.

\section{Conservation of $Q$ in The Branch Processes of Knotlike Cosmic
Strings}

In our previous work \cite{jiangying} it has been pointed out that, during
the evolution of cosmic strings, when the regular condition (\ref{regulcond}%
) fails, the branch processes (i.e. the splitting, mergence and
intersection) will occur; and in these branch processes, the sum of the
topological charges of all the final cosmic string(s) is equal to that of
all the initial cosmic string(s) at the bifurcation point, namely:

(a) for the case that one string $L$ split into two strings $L_1$ and $L_2,$
we have $W_L=W_{L_1}+W_{L_2};$

(b) the case that two strings $L_1$ and $L_2$ merge into one string $L:$ $%
W_{L_1}+W_{L_2}=W_L;$

(c) the case that two strings $L_1$ and $L_2$ meet, and then depart as two
other strings $L_3$ and $L_4:$ $W_{L_1}+W_{L_2}=W_{L_3}+W_{L_4}.$

In following we will show that when the branch processes of knotlike strings
occur, the topological invariant $Q$ of (\ref{CSact3}) [i.e. (\ref{CSact9})]
is preserved:

(i) The splitting case. We will consider one knot $\gamma $ split into two
knots $\gamma _1$ and $\gamma _2$ which are of the same self-linking number
as $\gamma \;$[$SL(\gamma )=SL(\gamma _1)=SL(\gamma _2)$], and then we will
compare the two numbers $Q_\gamma $ and $Q_{\gamma _1+\gamma _2}$ [where $%
Q_\gamma $ is the contribution of $\gamma $ to $Q$ before splitting, and $%
Q_{\gamma _1+\gamma _2}$ is the total contribution of $\gamma _1$ and $%
\gamma _2$ to $Q$ after splitting]. First, from the above text we have $%
W_\gamma =W_{\gamma _1}+W_{\gamma _2}$ in the splitting process. Second, on
the one hand, in the neighborhood of bifurcation point $(\vec{x}%
^{*},\;t^{*}) $, $\gamma _1$ and $\gamma _2$ are infinitesimally displaced
from each other; on the other hand, for a knot $\gamma $ its self-linking
number $SL(\gamma )$ is defined as 
\begin{equation}
SL(\gamma )=Lk(\gamma ,\;\gamma _V),
\end{equation}
where $\gamma _V$ is another knot obtained by infinitesimally displacing $%
\gamma $ in the normal direction $\vec{V}$ \cite{Witten}. Therefore 
\begin{equation}
SL(\gamma )=SL(\gamma _1)=SL(\gamma _2)=Lk(\gamma _1,\;\gamma _2)=Lk(\gamma
_2,\;\gamma _1),
\end{equation}
and 
\begin{equation}
Lk(\gamma ,\;\gamma _k^{\prime })=Lk(\gamma _1,\;\gamma _k^{\prime
})=Lk(\gamma _2,\;\gamma _k^{\prime })
\end{equation}
[where $\gamma _k^{\prime }$ denotes another arbitrary knot in the family ($%
\gamma _k^{\prime }\neq \gamma ,\;\gamma _k^{\prime }\neq \gamma _{1,2}$)].
Then, third, we can compare $Q_\gamma $ and $Q_{\gamma _1+\gamma _2}$ as:
before splitting, from (\ref{CSact9}) we have 
\begin{equation}
Q_\gamma =2\pi [W_\gamma ^2SL(\gamma )+\sum_{k=1\;(\gamma _k^{\prime }\neq
\gamma )}^N2W_\gamma W_{\gamma _k^{\prime }}Lk(\gamma ,\gamma _k^{\prime })],
\label{before}
\end{equation}
where $Lk(\gamma ,\gamma _k^{\prime })=Lk(\gamma _k^{\prime },\gamma )$;
after splitting, 
\begin{eqnarray}
Q_{\gamma _1+\gamma _2} &=&2\pi [W_{\gamma _1}^2SL(\gamma _1)+W_{\gamma
_2}^2SL(\gamma _2)+2W_{\gamma _1}W_{\gamma _2}Lk(\gamma _1,\gamma _2) 
\nonumber \\
&&+\sum_{k=1\;(\gamma _k^{\prime }\neq \gamma _{1,2})}^N2W_{\gamma
_1}W_{\gamma _k^{\prime }}Lk(\gamma _1,\gamma _k^{\prime
})+\sum_{k=1\;(\gamma _k^{\prime }\neq \gamma _{1,2})}^N2W_{\gamma
_2}W_{\gamma _k^{\prime }}Lk(\gamma _2,\gamma _k^{\prime })].  \label{after}
\end{eqnarray}
Comparing (\ref{before}) and (\ref{after}), we just have 
\begin{equation}
Q_\gamma =Q_{\gamma _1+\gamma _2}.
\end{equation}
This means that in the splitting process $Q$ is preserved.

(ii) The mergence case. We consider two knots $\gamma _1$ and $\gamma _2,$
which are of the same self-linking number, merge into one knot $\gamma $
which is of the same self-linking number as $\gamma _1$ and $\gamma _2$.
This is obviously the inverse process of the above splitting case, therefore
we have 
\begin{equation}
Q_{\gamma _1+\gamma _2}=Q_\gamma .
\end{equation}

(iii) The intersection case. This case is related to the collision of two
knots \cite{collision}. We consider two knots $\gamma _1$ and $\gamma _2,$
which are of the same self-linking number, meet, and then depart as other
two knots $\gamma _3$ and $\gamma _4$ which are of the same self-linking
number as $\gamma _1$ and $\gamma _2$. This process can be identified to two
sub-processes: $\gamma _1$ and $\gamma _2$ merge into one knot $\gamma $,
and then $\gamma $ split into $\gamma _3$ and $\gamma _4.$ Thus from the
above two cases (ii) and (i) we have 
\begin{equation}
Q_{\gamma _1+\gamma _2}=Q_{\gamma _3+\gamma _4}.
\end{equation}

Therefore we obtain the result that, in the branch processes during the
evolution of knotlike cosmic strings (i.e., the splitting, mergence and
intersection), the topological invariant $Q$ is preserved.

\section{Conclusion}

In this paper, the complex scalar quintessence field on the Riemann-Cartan
manifold of the early universe is considered. In Sect.II, it is revealed
that from the $U(1)$ field tensor $F_{\mu \nu }$ one can derive the cosmic
string structures, which just originate from the zero points of quintessence
field $\psi (x)$. In this section the Nielsen Lagrangian and Nambu action of
these strings are also simply discussed. In Sect.III, we emphasize on the
knotlike configurations in these cosmic strings. Based on the integral of
Chern-Simons $3$-form, we construct a topological invariant $Q$ for the
knotlike strings. It is pointed out that $Q$ is just the total sum of all
the self-linking and linking numbers of the knots family. In Sect.IV, it is
shown that $Q$ is preserved in the branch processes (i.e. the splitting,
mergence and intersection) during the evolution of knotlike cosmic strings.

At last there are two points which should be addressed. First, in this paper
we treat the cosmic strings as mathematical lines, i.e., the width of a
string is zero. This description is obtained in the approximation that the
radius of curvatures of a string is much larger than the width of the string 
\cite{Nielsen}. The further research on the width of cores of strings will
be detailed in our later papers. Second, in this paper the Nambu action of
cosmic strings has been simply discussed. Furthermore, the research of the
Higgs Lagrangian (i.e. the Ginzburg-Landau free energy) of knots as well as
the classification of knots will be detailed in our further work.

\section{Acknowledgment}

One of the authors XL is indebted to Dr. J. D. Bekenstein, Dr. A. A.
Kozhevnikov and Dr. R. Schiappa for their helpful advices and the
recommendation of their own outstanding work concerning the knotlike cosmic
strings. XL also would like to thank Dr. J. R. Ren and Dr. P. M. Zhang for
the instructive discussions and help.

This work was supported by the National Natural Science Foundation and the
Doctor Education Fund of Educational Department of the People's Republic of
China.

\end{document}